\documentclass[twocolumn,showpacs,
amsmath,amssymb,aps,prb,
]{revtex4-1}
\usepackage{graphicx}
\usepackage{dcolumn}
\usepackage{bm}

\usepackage{mathtools}
\DeclarePairedDelimiter{\ceil}{\lceil}{\rceil}

\begin{document}

\preprint{APS/123-QED}

\title{Non-equilibrium thermal transport in the quantum Ising chain}

\author{Andrea De Luca}
\email{andrea.deluca@lptens.fr}
\affiliation{Laboratoire de Physique Th\'eorique de l'ENS \& Institut de Physique Theorique Philippe Meyer, France.}
\author{Jacopo Viti}
\email{jacopo.viti@lptens.fr}
\affiliation{Laboratoire de Physique Th\'eorique de l'ENS, CNRS \& Ecole Normale Sup\'erieure de Paris, France.}
\author{Denis Bernard}
\email{denis.bernard@lptens.fr}
\affiliation{Laboratoire de Physique Th\'eorique de l'ENS, CNRS \& Ecole Normale Sup\'erieure de Paris, France.}
\author{Benjamin Doyon}
\email{benjamin.doyon@kcl.ac.uk}
\affiliation{Department of Mathematics, King's College London, United Kingdom.}


\date{\today}

\begin{abstract}
We consider two quantum Ising chains initially prepared at thermal equilibrium but with different temperatures 
and coupled at a given time through one of their end points. In the long-time limit the system reaches a non-equilibrium steady state.
We discuss properties of this non-equilibrium steady state,  and
characterize the convergence to the steady regime.
We compute the mean energy flux through the chain and show that the heat transport is ballistic. We derive also the
large deviation function for the quantum and thermal fluctuations of this energy transfer. 

\end{abstract}

\maketitle

\newcommand{\beq}{\begin{equation}}
\newcommand{\eeq}{\end{equation}}
\newcommand{\Ord}[1]{{O}\left(#1\right)}

\newcommand{\sign}{\,{\rm sign}}
\newcommand{\arctanh}{\,{\rm arctanh}}

\def\bra#1{\mathinner{\langle{#1}|}} 
\def\ket#1{\mathinner{|{#1}\rangle}} 
\newcommand{\braket}[2]{\langle #1|#2\rangle} 
\def\Bra#1{\left<#1\right|} 
\def\Ket#1{\left|#1\right>}
\providecommand{\abs}[1]{\lvert#1\rvert}  
\providecommand{\norm}[1]{\lVert#1\rVert}

\newcommand{\twovec}[2]{\left[\begin{array}{c}
#1\\#2
\end{array}
\right]}

\newcommand{\avg}[2]{\langle #1 \rangle_{\mbox{\tiny #2}}}
\def\Tr{\operatorname{Tr}}
\def\diag{\operatorname{diag}}

\def\frl{\phi_{r/l}}
\def\fr{\phi_{r}}
\def\fl{\phi_{l}}
\def\fRL{\psi_{R/L}}
\def\fR{\psi_{R}}
\def\fL{\psi_{L}}

\def\irl{\Phi_{r/l}}
\def\ir{\Phi_{r}}
\def\il{\Phi_{l}}
\def\iRL{\Psi_{R/L}}
\def\iLR{\Psi_{L/R}}
\def\tiRL{\tilde\Psi_{R/L}}
\def\iR{\Psi_{R}}
\def\iL{\Psi_{L}}
\newcommand{\ialpha}[1]{\Psi_{#1}}

\def\fLW{\omega}
\def\fFO{\Psi}
\def\fFW{\psi}
\def\fFOstat{\vec{\fFO}^{\mbox{\tiny stat}}}
\def\FRL{\vec{\Upsilon}}
\def\ttheta{\tilde\theta}

\def\rightP{\mathcal{P}}
\def\mysmall{\delta}

\def\MM{M}
\def\bigM{\mathcal{M}}
\def\bigN{\mathcal{N}}
\def\bigc{\mathbf{c}}
\def\bigQ{\mathcal{Q}}
\def\bigpsi{\mathbf{\Psi}}
\def\bigphi{\mathbf{v}}
\def\bigA{\mathcal{A}}
\def\bigB{\mathcal{B}}
\def\bigAt{\tilde{\mathcal{A}}}
\def\smallM{m}
\def\Hleft{H_l}
\def\Hright{H_r}
\def\smallh{h}
\def\betaleft{\beta_l}
\def\betaright{\beta_r}
\newcommand\pv[1]{P_v\left[#1\right]}
\def\mom{p}
\def\Li{\operatorname{Li}}
\def\rhostat{\rho_{\text{\tiny stat}}}
\graphicspath{ {../} }

\section{Introduction}
Recent experiments \cite{blanter2000shot} have strongly renewed the attention on non-equilibrium quantum dynamics.
In particular, it has become possible to accurately measure the current \cite{bylander2005current} flowing between two leads driven
out of equilibrium through macroscopical
control parameter, e.g. temperature or voltage gradient. Moreover, several proposals have been advanced in order to theoretically characterize and measure the quantum fluctuations of the current \cite{gustavsson2006counting, bomze2005measurement, levitov1993charge, esposito2009nonequilibrium}.
In this framework, 1d systems are peculiar: on one side they represent an approximate description for 3d systems with
strong anisotropy; on the other side, the dynamics is anomalous because of the role of purely elastic scattering processes.
In particular, in some remarkable cases, the low-energy description
can be given in terms of integrable field theories \cite{zamolodchikov1979factorized, mussardo2009statistical}. In these cases, the presence of an infinite set of conserved charges may result in a non-vanishing Drude weight and a ballistic heat transport\cite{sirker2009diffusion}.
It is then crucial to establish how universal\cite{karrasch2012nonequilibrium} these properties are and what the role played by dimensionality is.
In this paper, we focus on the analysis of the energy-current steady state in the simplest example of solvable lattice models,
the quantum Ising chain. The protocol that we choose for constructing the out-of-equilibrium steady state is that of Hamiltonian reservoirs\cite{spohn1977stationary, spohn2007irreversible}, defined as follows.
At time $t = 0$, the system is prepared as two independent, finite, bounded (left and right) chains thermalized at different temperatures $\beta_l$ and $\beta_r$. The initial system density matrix is $\rho_0 = \rho_l \otimes \rho_r $, where $\rho_{r/l} = Z_{r/l}^{-1}\exp(-\beta_{r/l} H_{r/l})$.
Then the two chains are coupled at their end-points in the middle, producing a single homogeneous bounded chain twice as long with Hamiltonian $H$, and the system is let evolve unitarily. We will study and characterize the infinite-chain, long-time
dynamics of local observables, defining the stationary density matrix $\rhostat$. Although the evolution of the whole system is unitary, as long as the evolution time remains smaller than the time an elementary excitation
in the bulk needs to reach the boundary,
any finite part of the chain around the connection point is effectively open and coupled to two asymptotic thermal baths. The stationary density matrix $\rhostat$ describes observables on such finite parts of the chain, not in the asymptotic baths.
The stationary state density matrix $\rhostat$ (given in \eqref{nedm} below) shows
a factorized form in terms of its chiral components. Furthermore,  convergence to the steady state is non-universal
and follows a $t^{-1/2}$ behavior. We will characterize the heat transport along the chain,
computing the mean current $\mathcal{J}$, see \eqref{current}, and its
large-deviation function $F(\lambda)$ which satisfies fluctuation-dissipation relations, see \eqref{large_deviation_final}.
As already remarked in \cite{bernard2012energy} the derivative of $F(\lambda)$ can be obtained
from the knowledge of the mean current at shifted values of the temperatures $\beta_l$ and $\beta_r$, 
see \eqref{derivativeJ}.

The density matrix $\rhostat$ was already obtained in the anisotropic XY chain (a family of models which includes the Ising chain)\cite{aschbacher2003non}
within the formalism of $C^*$ algebra, where the factorized form was observed; however the approach to the steady state and the full counting statistics were not discussed. It was also obtained in 1+1-dimensional conformal field theory (CFT)\cite{bernard2012energy,bernard2013non} and massive quantum field theory (QFT)\cite{doyon2012nonequilibrium}, where a similar factorized form was observed.
Note that an other approach to treat $1d$ open quantum systems based on Lindblad dynamics has
been developed\cite{prosen2011open}.

This paper is organized as follows: in Sec. \ref{solution}, we present a brief description of the diagonalization procedure and the thermodynamic limit (further details are given in the Appendix); 
in Sec. \ref{ness}, we show that local observables have a well defined large time limit that implies
a factorized form for $\rhostat$; 
in Sec. \ref{sec_current}, we derive an explicit expression for the stationary energy flow; 
in Sec. \ref{sec_ldf} the same formalism is applied to obtain the large deviation function.

\section{Exact solution
\label{solution}
}
The Ising Hamiltonians for the right and left chains are defined as, respectively ($J>0$)
\begin{align}
\label{righthamiltonian}
H_r &= -\frac J 2 \left[ \sum_{i=1}^{N-1} \sigma_i^x \sigma_{i+1}^x + h \sum_{i=1}^N \sigma_i^z \right]\\
\label{lefthamiltonian}
H_l &= -\frac J 2 \left[ \sum_{i=1}^{N-1} \sigma_{-i}^x \sigma_{-i+1}^x + h \sum_{i=0}^{N-1} \sigma_{-i}^z \right]\; .
\end{align}
These spin chains can be exactly solved employing the usual Jordan-Wigner procedure \cite{lieb1961two}. After having introduced
the canonical fermionic operators 
\beq
\label{JW}
c_i=-\bigl(\prod_{j<i}\sigma_j^{z}\bigr)\sigma_i^{-}
\eeq
and their hermitian conjugates, where $\sigma_{i}^{\pm}=1/2(\sigma_i^{x}\pm i\sigma_{i}^{y})$, the Hamiltonians (\ref{righthamiltonian}, \ref{lefthamiltonian}) become quadratic forms that can be diagonalized 
by means of a Bogoliubov transformation (see the Appendix), using new canonical operators
\beq
\label{Bogoliubov}
\frl(k) = \sum_{i = -N+1}^N \bigl[\omega_{r/l}^i(k) c_i + \xi_{r/l}^i(k) c_i^\dag\bigr]   
\eeq
with $k = {1,\ldots,N}$. The Hamiltonians take the form
\beq
H_{r/l} =  \sum_{k=1}^N \epsilon(\theta_k) \frl^\dag(k) \frl(k);
\eeq
the one-particle spectrum is  $\epsilon(\theta) =J\sqrt{h^2 + 1 - 2h \cos \theta}$ and the angles $\theta_k$ 
are solutions of the transcendental equation
\beq
\label{thetatrasc}
\frac{\sin [(N+1) \theta_k] }{\sin (N \theta_k)} = h^{-1} \;.
\eeq
We restrict to the case $h>1$, i.e. the paramagnetic phase, where the equation \eqref{thetatrasc} has exactly $N$ real solutions
in the interval $[0,\pi]$. 
In an analogous way, one can treat the full chain Hamiltonian $H = H_l + H_r + V$ where $V = -\frac J 2 \sigma_0^x \sigma_1^x$.

We now consider the thermodynamic limit $N\to \infty$. For the right / left chain with Hamiltonian $H_{r/l}$ the result is a half-infinite chain, either extending to the right or to the left.
The solutions of the equation \eqref{thetatrasc} distribute uniformly in the interval $[0,\pi]$.
Therefore the variable $\theta$ becomes continuous,
and we can define properly normalized operators $\irl$ that satisfy
\footnote{Formally the limit can be written as
$$ \irl (\theta) = \lim_{N\to \infty} \sqrt\frac{N}{\pi} \frl (k_N(\theta)) $$
and $k_N(\theta)$ can be  chosen as $\ceil{\frac{\theta N}{\pi}}$.
}  standard anticommutation relations
\beq
\{ \irl^\dag (\theta), \irl(\theta')\} = \delta(\theta- \theta').
\eeq
For $H$ the result is an infinite chain. There a bit of care is needed. In the $N\to \infty$ limit, the chain, already symmetric under parity, becomes also translation invariant. Hence the translation operator $\rightP$, defined by $\rightP^\dag \sigma_{i}^{\alpha} \rightP = \sigma_{i-1}^{\alpha} $ for $\alpha = x,y,z$, can be diagonalized simultaneously with $H$. Since $\rightP\leftrightarrow \rightP^\dag$ under parity, this imposes a two-fold degeneracy of the single-particle spectrum, which can be seen explicitly checking the odd and even $k$ solutions for large $N$ of \eqref{thetatrasc}.
We stress that, in the thermodynamic limit, $H_0$ and $H$ have, as expected, the same single-particle spectrum with the same degeneracy.
In the single-particle degenerate eigensubspace of $H$, it is possible to fix the two fermionic operators at each $\theta$
by imposing that they have definite $\rightP$-eigenvalues, which turn out to be $e^{\pm i\theta}$. We then define canonical operators for the full chain
\beq
\{ \iRL^\dag (\theta), \iRL(\theta')\} = \delta(\theta- \theta')
\eeq
with
\beq
\label{RLmovers}
\rightP^\dag \iRL(\theta) \rightP= e^{\mp i\theta}\iRL(\theta).
\eeq
The Hamiltonian takes the form
\beq
\label{hamTL}
H = \int_0^\pi d\theta \, \epsilon(\theta) [\iR^\dag(\theta)\iR(\theta) +
\iL^\dag(\theta)\iL(\theta)] \equiv H_R + H_L\;.
\eeq
For the lattice definition of $\iR$ and $\iL$ we refer to the Appendix.
\section{Non-equilibrium steady state
\label{ness}
}
\subsection{Factorization of the stationary density matrix}
We introduce density matrices 
as functionals on the space of observables. We will argue that for local observables, 
i.e. the class of operators with support localised close to the contact point and of finite extent,
the following limit exists
\beq
\label{rhostatdef}
\lim_{t\to\infty}\lim_{N\to\infty} \Tr [O(t) \rho_0] = \Tr [O \rhostat] \;.
\eeq
thus defining the action of $\rhostat$ \cite{aschbacher2003non, ruelle2000natural, aschbacher2006topics}.
It is natural to reformulate it introducing the $S$-matrix, defined as
\beq
\label{smatrix}
S = \lim_{t \to \infty} e^{-i H t} e^{i H_0 t},
\eeq
by which we can formally write $\rhostat = S\rho_0 S^\dag$. This has to be intended 
as $\Tr(O S\rho_0 S^\dag) = \Tr(S^\dag O S\rho_0)$, for a local observable $O$, which is nothing more than \eqref{rhostatdef}. The action of $S$ over observables can then be obtained 
once its action over the fermionic operators $\iRL(\theta)$ is known.

It is useful to observe that if the initial state is at equilibrium, $\beta_l = \beta_r$, it is natural to expect thermalization with the final Hamiltonian, i.e. in this case $ \rhostat = Z^{-1} e^{-\beta H} = Z^{-1} S e^{-\beta H_0} S^\dag$.
It follows that one can formally write, inside traces $S H_0 S^\dag = H$, namely $S$ intertwines between the two Hamiltonians, as expected on the general basis of the Gell-Mann-Low theorem \cite{gell1951bound}.
This property is useful because it suggests that $S^\dag \iRL(\theta) S$ is a linear combination of $\irl(\theta)$ and can therefore be represented as a $2\times2$ orthogonal matrix
\beq
S^\dag \twovec{\iR(\theta)}{\iL(\theta)} S = \left(
\begin{array}{cc}
 \cos a(\theta) & \sin a(\theta) \\
 \sin a(\theta) & -	\cos a(\theta)
\end{array}
\right)
\twovec{\ir(\theta)}{\il(\theta)} \;.
\eeq
We verified explicitly the validity of this claim.
We remark once more that the action of the $S$ matrix
is well defined only on local operators. Therefore $\iRL(\theta)$ must be replaced by a 
smooth superposition of fermionic operators\cite{weinberg2005quantum}, which is local in the lattice. Such superpositions include both left and right movers:
\beq	
\label{wavepacket}
\tilde\Psi(\theta) = \int_{-\pi}^0 d\theta' \,g_{\theta}(\theta') \iL(-\theta') +
\int_0^\pi d\theta'\,g_\theta(\theta') \iR(\theta')\;.
\eeq
Here $g_{\theta}(\theta')$ is a wave-packet centered in $\theta$. Locality, more precisely the exponential vanishing of $\{\tilde\Psi(\theta),c_j\}$,  $\{\tilde\Psi(\theta),c_j^\dag\}$ at large $|j|$, imposes only that $g_{\theta}(\theta')$ be $2\pi$-periodic in $\theta'$ and analytic on a neighborhood of the real $\theta'$-line. If the wave packet is peaked enough only one of the two terms essentially contribute in (\ref{wavepacket}). At the end of the calculation, we will need to take the delocalization limit $g_\theta(\theta')\to \delta(\theta-\theta')$.

We first provide an heuristic argument to specify the value of $a(\theta)$, that we will prove explicitly later. This heuristic argument parallels the more precise derivations developed in CFT\cite{bernard2012energy,bernard2013non} and in massive QFT\cite{doyon2012nonequilibrium}.
The eigenfunctions of the single-particle Hamiltonians for the left and right half chains can be thought of
as stationary waves due to the boundary condition at the origin. 
Consider an operator which is a superposition of $\Phi_l$ and which creates a perturbation localized on the left, say around the site $j<0$. The time evolution with $H_0$ will split it into two perturbations formed by right and left movers, whose centers move respectively towards the left and the right.
However, that moving towards the right will eventually
hit the right-boundary thus inverting its motion under the $H_0$ dynamics. Therefore,
for large time $t$, the evolved operator will resemble an operator formed of right movers evolved with the Hamiltonian $H$. Thus we expect 
\beq
\label{factorizationnaive}
e^{i H_0 t} \irl(\theta)e^{-i H_0 t}~\stackrel{t\to\infty}{=}~e^{i H t} \iLR(\theta)e^{-i H t}
\eeq
which would correspond to $a(\theta) = \pi/2 $.

We now give a more formal derivation. Inverting the relations between the fermionic fields and the local fermions $c_i$'s and $c_i^{\dagger}$'s,
one obtains the change of basis matrices
\beq
\hspace{-0.2cm}\twovec{\iR(\theta)}{\iL(\theta)} = \int_0^\pi \hspace{-0.2cm}d\theta' \smallM(\theta, \theta') \twovec{\ir(\theta')}{\il(\theta')} + \tilde \smallM(\theta, \theta') \twovec{\il^\dag(\theta')}{\il^\dag(\theta')} 
\eeq
If we take a wave packet as in \eqref{wavepacket} and we act upon it with the
$S$ matrix we must deal with the expression
\beq
\label{largetlimit}
\hspace{-0.2cm}\lim_{t \to \infty} \int_0^\pi \hspace{-0.2cm} d\theta' \int_0^\pi \hspace{-0.2cm} d\theta''g_{\theta}(\theta')e^{-i t(\varepsilon(\theta') - \varepsilon(\theta''))} \smallM(\theta',\theta'') \twovec{\ir(\theta'')}{\il(\theta'')} 
\eeq
and a similar one with $g_\theta(\theta')\mapsto g_\theta(-\theta')$, and for $\tilde m(\theta',\theta'')$.
It is useful to change variable in the integral in $\theta'$, by introducing $z = e^{- i \theta'}$.
\begin{figure}[ht]
\centering
\includegraphics[width=0.8\columnwidth]{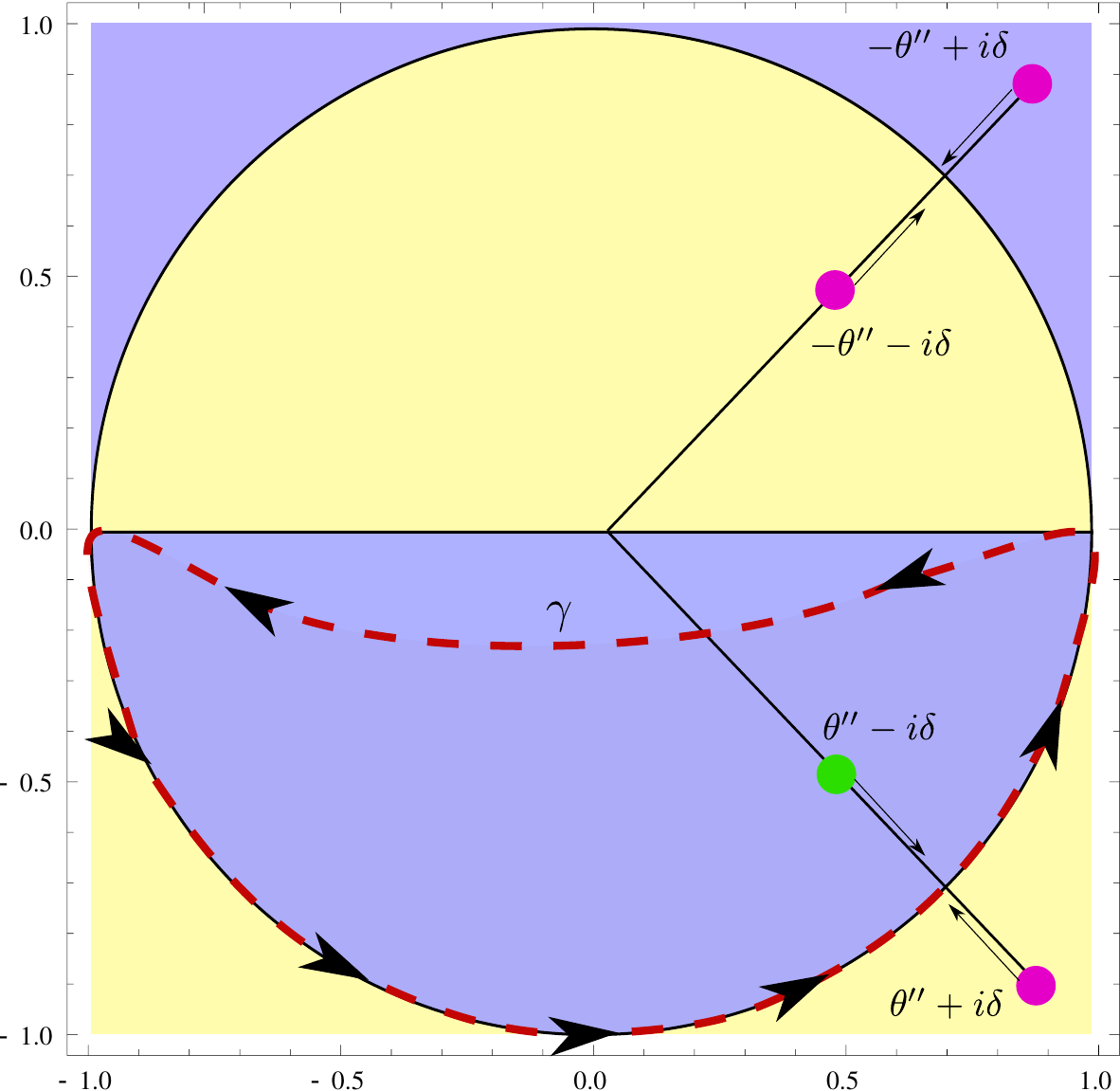}
\caption{Position of the poles $m (\theta', \theta'')$ in the $z = e^{-i\theta'}$ plane; 
the colors refer to the sign of the imaginary part of $\epsilon(\theta')$: positive for yellow and negative for blue. The integral over $\theta'$
becomes the integral over the lower semi-circumference: we close the contour inside the lower semicircle, where $\Im \varepsilon(\theta') < 0$ and such contribution vanishes in the $t \to \infty$ limit.} 
\label{contourplot}
\end{figure}
In the $z$ variable, one has to integrate over the lower semi-circumference as shown in Fig.~\ref{contourplot}, where $g_\theta$ is analytic. We may also assume $g_\theta$ to be analytic on the interval $[-1,1]$. Then, it is possible to close the contour with a curve $\gamma$ in the lower semicircle, where the imaginary part of $\varepsilon(\theta')$ is negative; this does not affect the result in the large-$t$ limit as the integral over $\gamma$ vanishes in this limit. The integral reduces to the sum of the residues inside this region. The relevant contribution coming from the change of basis matrices is
\beq
m(\theta', \theta'') = \frac{1}{2\pi i}\left(\frac{1}{\theta' - \theta'' + i
\delta} - c.c.\right)\sigma_x + \ldots
\eeq
where the remaining part and $\tilde m(\theta',\theta'')$ are not singular
inside the integration domain and $\sigma_x$ is a Pauli matrix. The
regularization $\delta\sim 1/N>0$ is associated with the finite-volume regularization.

According to
\eqref{rhostatdef}, the thermodynamic limit has to be taken before the large
time limit. Further, the large-time limit should be taken before the delocalization limit. Sending $\delta$ to $0^+$ before $t\to \infty$, in the large $t$ limit all the poles inside the contour have vanishing
residue, except for the one at $\theta' = \theta'' - i\delta$. After this, the wave-packet
width can be safely sent to $0$, i.e. $g_\theta (\theta') = \delta(\theta -
\theta')$ and one obtains the result anticipated in \eqref{factorizationnaive}.
In particular, the non-equilibrium density
matrix  takes the factorized \cite{aschbacher2003non} form
\beq
\label{nedm}
\rhostat = Z^{-1} e^{-\beta_r H_L} \otimes e^{-\beta_l H_R} \; .
\eeq
This expression can be  check to be in the form proposed in \cite{hershfield1993reformulation}
for a non-equilibrium density matrix.

Notice that the opposite limit $t\to -\infty$ can be treated analogously, but
the contour in Fig.~\ref{contourplot} has to be closed outside the unit circle
in the lower half-plane. The relevant pole becomes $\theta' = \theta'' + i
\delta$ and $a(\theta) = 0$. This reflects the fact that time-reversal symmetry
is actually broken in the steady state, but $\mathcal{PT}$-symmetry is still
preserved.
\subsection{Approach to the steady-state}
After having investigated the stationary density matrix, it can be useful to
study how fast the convergence is. As we saw, all the poles inside the contour
correspond to exponentially vanishing corrections. It remains to estimate
the contribution of the integral over the curve $\gamma$ and we employ the
saddle-point approximation. We observe that, however we choose $\gamma$ inside
the lower semicircle, the largest values of the imaginary part of
$\epsilon(\theta')$ will be at the boundaries, i.e. $\theta'^\star
=\{ 0,\pi\}$. We can therefore expand all the functions close to them obtaining,
as usual, gaussian integrals that can be readily performed. This provides a
power-law approach $\simeq t^{-1/2}$. This correction is due to the
zero modes in the single particle spectrum, that do not move during the
dynamics, giving place to a field localized in the original center of the wave
packet. Even in the $h \to 1$ gapless limit, this contribution is still present
due to the zero mode at $\theta' = \pi$, at the end of the spectrum, showing
that it can not be deduced from the low-energy physics. A $t^{-1/2}$ power law approach to the steady state was also observed in the context of quantum quenches in the Ising model\cite{calabrese2011quantum}.
\section{Heat current
\label{sec_current}
}
In this section we will derive the stationary energy flow between the two halves of the chain. We introduce $E=\frac{1}{2}(H_r-H_l)$ whose variation is the total amount of energy transferred from
the left to the right part of the chain. Its time derivative $p=i[H,E]$ is local with
support on the lattice sites $j=0,1$, and one can therefore expect its long-time limit expectation value to converge to a stationary value
\beq
\label{currentdef}
\lim_{t\rightarrow\infty}\Tr[p(t)\rho_0]=\Tr[p\rhostat]\equiv \mathcal{J}.
\eeq
In the stationary state the mean energy current stays constant to the value $\mathcal{J}$, reached in the long-time limit. 
In terms of local fermions $c_j$'s and $c_j^{\dagger}$'s, $p=\frac{i hJ^2}{2} (c_0^\dag c_1 - c_1^\dag c_0)$. When expressed
through the non-local left and right mover operators \eqref{RLmovers} the operator $p$ is represented by the quadratic form
\beq
\label{p_local}
p=J^2\int_{0}^{\pi}d\theta d\theta'~\left[\begin{array}{cc}\iR^{\dagger}(\theta)\iL^{\dagger}(\theta)\end{array}\right]
\mathcal{A}(\theta,\theta')\twovec{\iR(\theta')}{\iL(\theta')},
\eeq
where the explicit expression for the $2\times 2$ matrix $\mathcal{A}(\theta,\theta')$ is not needed here. The operator $p$ can be seen as a momentum density at site 0. The momentum density at site $j$ is
$p_j=\mathcal{P}^jp~[\mathcal{P}^{\dagger}]^j$, and summing over $j\in\mathbb Z$ one obtains the total momentum $P$ of the chain, which satisfies $[P,c_j] = \frac{ihJ}2 (c_{j-1}-c_{j+1})$. Hence\footnote{$P$ is a diagonal quadratic form on the basis of the left and right movers and we only have to identify 
its eigenvalues. Since the full chain is invariant under discrete translations $P$ is also diagonalized by the modes $a(\theta)$
such that
$$c_j= \int_{-\pi}^{\pi}\frac{d\theta}{\sqrt{2\pi}} e^{ij\theta}a(\theta).$$
Substituting this expression into the form $P=\frac{1}{J}\sum_{j\in\mathbb Z} p_j$ leads immediately to
\eqref{momentum}.},
\beq
\label{momentum}
P=hJ\int_{0}^{\pi}d\theta\sin\theta~[n_R(\theta)-n_L(\theta)]
\eeq
with $n_{R/L}(\theta)$ the number operators for the right/left moving fermions. 
Applying now the definitions \eqref{RLmovers} and summing over $j$ in (\ref{p_local}) one concludes that the diagonal elements
of the matrix $\mathcal{A}$ are fixed to be $\mathcal{A}(\theta,\theta)={\rm diag}(\pm\frac{h}{2\pi}\sin\theta)$. It turns out that this
is sufficient for our purposes. It is indeed clear that the expectation value computed with the stationary matrix $\rhostat$ of a bilinear fermionic operator
satisfies
\beq
\label{diagexpval}
\Tr[\ialpha{\alpha}^{\dagger}(\theta)\ialpha{\beta}(\theta')\rhostat]=\delta_{\alpha,\beta}\delta(\theta-\theta')\frac{1}
{e^{\beta_\alpha\varepsilon(\theta)}+1},
\eeq
for $\alpha, \beta=R,L$, with $\beta_R\equiv\beta_l$ and $\beta_L\equiv\beta_r$. Performing the integration over the variable $\theta$ we finally derive
\begin{align}
\label{current}
 \mathcal{J}(\beta_l,\beta_r)=&\frac{1}{2\pi\beta_l^2}\bigl[j\bigl(
 \beta_lJ(h+1)\bigr)-j\bigl(\beta_lJ(h-1)\bigr)+\nonumber\\
 &-\frac{1}{2\pi\beta_r^2}[j\bigl(\beta_rJ(h+1)\bigr)-j
 \bigl(\beta_rJ(h-1)\bigr)\bigr],
\end{align}
where  $j(x)=\Li_2(-e^{-x})
 -x\log(1+e^{-x})$.
 \begin{figure}[ht]
\centering
\includegraphics[width=0.8\columnwidth]{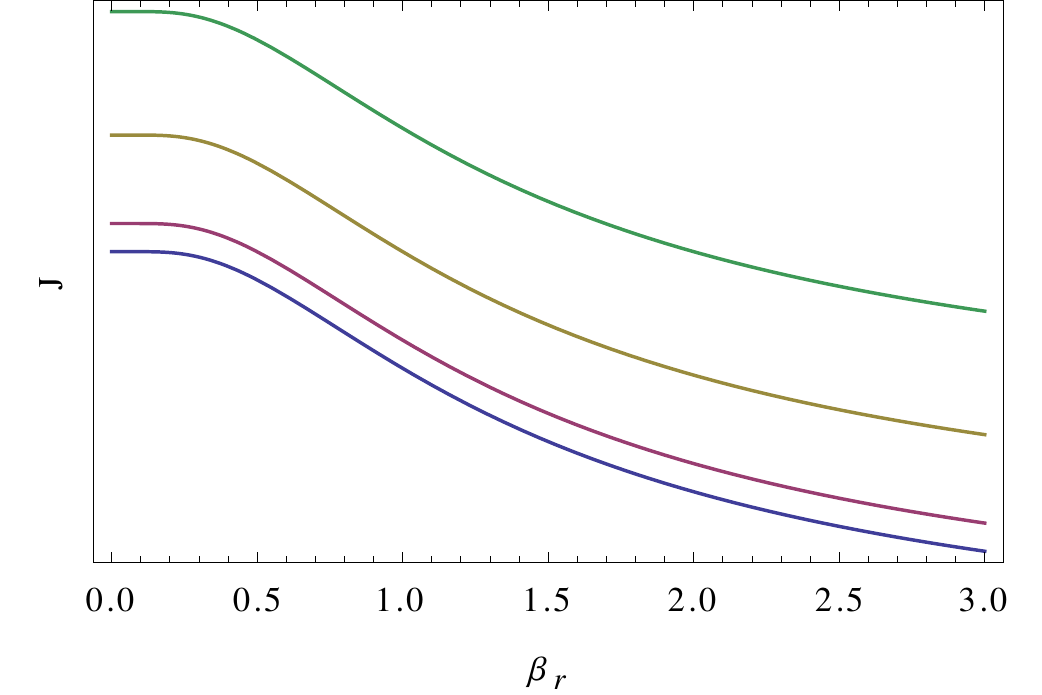}
\caption{The current $\mathcal{J}$ vs $\beta_r$ for different $\beta_l=  \{0.1, 0.5, 1, 2.0\}$ (from bottom upwards) at $h=2$ . Due to factorization, different curves are just translated along the $y$-axis. Experimentally one just needs to measure the current for two different left/right temperatures to verify whether factorization occurs, i.e. $\partial_{\beta_l} \partial_{\beta_r} \mathcal{J} = 0$. 
} 
\label{currenttranslation}
\end{figure}
The expression \eqref{current} can be evaluated at the critical point $h=1$. At the critical point, its universal conformal behavior as $\beta_{l/r}\gg J^{-1}$ is obtained using $\Li_2(-1)=-\pi^2/12$, $\Li_2(0)=0$, giving $\frac{\pi}{24}(\beta_l^{-2}-\beta_r^{-2})$, 
in agreement with the general theory developed in \cite{bernard2012energy, bernard2013non}, for the value $1/2$ of the CFT central charge.
The universal scaling limit leading to a massive QFT can be explicitly read off in \eqref{current},
taking $h\to 1$ and $\beta_{l,r}J \to \infty$ with fixed products $\beta_{l,r}J(h-1)$. Plots of $\mathcal{J}$
as a function of $\beta_r$ and for different values of $\beta_l$ are shown in Fig. ~\ref{currenttranslation}.
\section{Large deviation function
\label{sec_ldf}}
We have seen that, given the factorized form of the steady state, the average current is related to the asymmetry in the number of movers in the two directions. 
However a large amount of information is still hidden inside the higher moments. In order to define them, one introduces
the total amount of energy transferred within a time interval $[t_1, t_2]$ as 
\beq
\label{energytransferred}
\Delta E(t_2,t_1) \equiv \int_{t_1}^{t_2} p(s) ds \;.
\eeq
If $\mathbb P [\Delta E(t,0)=x]$ is the probability to measure a value $x$ for the observable $\Delta E(t,0)$, it is possible to define
its large deviation function 
\beq
\label{probflambda}
F(\lambda)=\lim_{t \to \infty} \frac{1}{t}\log \sum_{x} e^{-\lambda x}~ \mathbb P [\Delta E(t,0)=x] \,
\eeq
with $\lambda$ chosen to ensure convergence and the sum running over the spectrum of $\Delta E(t,0)$. In a quantum system, beside the formal definition \eqref{probflambda}
one must also specify the protocol adopted to measure the observable $\Delta E(t,0)$. This is particularly relevant in our case
because to measure $\Delta E (t,0)$ one needs to perform two projective measurements on the system at different
times and in general $[E(t),E(0)]\not=0$, see \cite{esposito2009nonequilibrium}. A possible definition, based on
an indirect measurement protocol, has been proposed in \cite{levitov1993charge} leading to
\beq
 \label{Levitov_def}
 \tilde F(\lambda) = \lim_{t \to \infty} \frac{1}{t}\log \Tr[\rhostat e^{-\lambda E(t)} e^{\lambda E(0)}]\,.
 \eeq
We will not deal explicitly with \eqref{Levitov_def}, but instead with the more manageable
\beq
\label{fullcountingdef}
F(\lambda) = \lim_{t \to \infty} \frac{1}{t}\log \Tr[\rhostat e^{-\lambda \Delta E(t,0)}]\; .
\eeq
Although this definition does not correspond to a realistic experimental setup, we will argue at the end
of the section that the large $t$-limit and the properties of the stationary state \eqref{nedm} are such that
\eqref{Levitov_def} and \eqref{fullcountingdef} coincide.

\begin{figure}[ht]
\centering
\includegraphics[width=0.8\columnwidth]{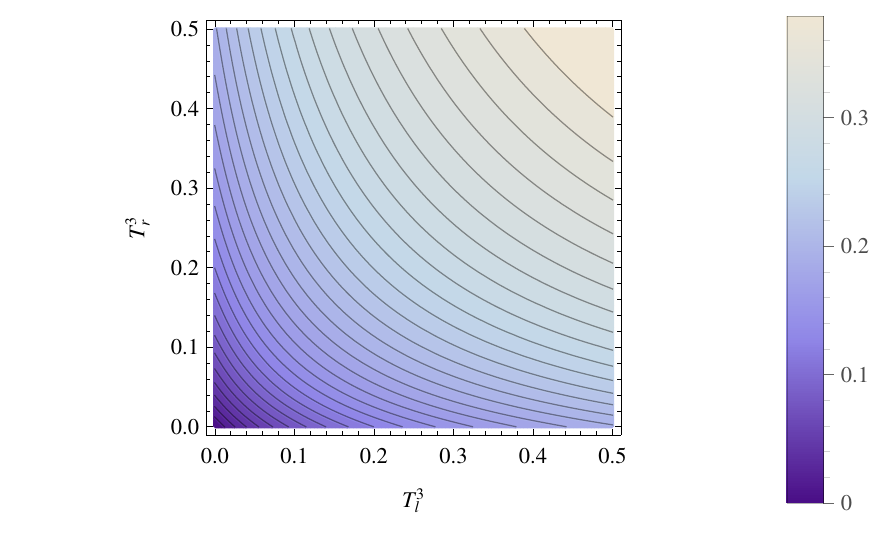}
\caption{Level-plot of the second moment $\Sigma_2 = \left.\partial_{\lambda}^2 F(\lambda)\right|_{\lambda=0}$ for $h=1$. In the conformal limit it takes the simple form $\Sigma_2 = \pi/12( \beta_l^{-3} + \beta_r^{-3})$. The scaling limit is therefore manifest in the plot as the straightening of the level-lines going at lower temperatures. The curvature is a lattice effect. Notice it is not vanishing on the diagonal, for $\beta_l = \beta_r$, as a result of thermal fluctuations.
} 
\label{contourplotfluctuation}
\end{figure}

The present derivation parallels that done in CFT\cite{bernard2013non}. By taking the derivative with respect to $\lambda$ of \eqref{fullcountingdef}
and using  time-translation invariance of  the steady-state one gets
\beq
\label{fullcountingder}
-\partial_{\lambda}F(\lambda)=\lim_{t\rightarrow\infty}\frac{1}{t}\frac{\Tr[\Delta_t E~
 e^{-\lambda \Delta_t E}\rhostat]}{\Tr[ e^{-\lambda \Delta_t E} \rhostat]},
\eeq
where $\Delta_{t} E \equiv \Delta E(t/2, -t/2)$. Substituting \eqref{p_local} in \eqref{energytransferred}, one can take the large $t$ limit
using the standard representation $\delta(x) = \lim_{t\to \infty} \sin(t x)/\pi x$, obtaining
\beq
\label{largetenergytransferred}
\lim_{t\to\infty} \Delta_t E = H_R - H_L \;. 
\eeq
In order to take the $t\to\infty$ limit, \eqref{largetenergytransferred} can be substituted in the exponent of \eqref{fullcountingder}. One realizes that the resulting expression
is equivalent to 
\beq
-\partial_{\lambda}F(\lambda)=\lim_{t\rightarrow\infty}\frac{1}{t}\int_{-t/2}^{t/2}ds~\Tr\bigl[p(s)\rhostat(\lambda)\bigr],
\eeq
where $\rhostat(\lambda)$ is the factorized density matrix in \eqref{rhostatdef} with inverse temperatures shifted as $\beta_l \to \beta_l+\lambda$ and $\beta_r \to \beta_r-\lambda$. 
The trace can be now computed using \eqref{p_local} and \eqref{diagexpval}. The integrand becomes $s$-independent, thus canceling the $1/t$ prefactor and leading to
\beq
\label{derivativeJ}
- \partial_\lambda F(\lambda) = \mathcal{J}(\beta_l+\lambda, \beta_r-\lambda) \;.
\eeq
Performing the integration over $\lambda$, one obtains
\beq
\label{large_deviation_final}
 F(\lambda)=[f_{\beta_l}(\lambda)-f_{\beta_l}(0)]+[f_{\beta_r}(-\lambda)-f_{\beta_r}(0)],
\eeq
where we have defined
\beq
f_{\beta}(x)=\frac{\Li_2\bigl(-e^{-(\beta+x)J(h+1)}\bigr)-\Li_2\bigl(-e^{-(\beta+x)J(h-1)}\bigr)}{2\pi(\beta+x)}.
\eeq
We remark that this expression is non-trivial also when $\beta_l = \beta_r$, where, although the current is vanishing,
higher moments are not, because of thermal fluctuations (see Fig.~\ref{contourplotfluctuation}).

As in \eqref{current}, the conformal and the scaling limit can be readily obtained. In particular, \eqref{derivativeJ} has been shown in \cite{bernard2012energy} to be valid generally at the gapless point, and is shown to be generally related to ${\cal PT}$-symmetry in \cite{bernard2013time}.

Finally, we notice that this expression is indeed consistent with the fluctuation-dissipation relations \cite{gallavotti1995dynamical, gallavotti1995dynamical1}
\beq
\label{gallavotti}
 F(\lambda) = F(\beta_r -\beta_l- \lambda) \; .
 \eeq
 
 Now, we come back to the problem of showing the equivalence between the two definitions \eqref{Levitov_def} and \eqref{fullcountingdef}.
 A rigorous proof seems to be non-trivial even in the simple model we are considering here. However, we provide a formal and heuristic argument
 for which they must coincide. A more thorough discussion has been recently published\cite{bernard2013time}. From time-translation invariance inside the steady state and applying
 the $S$-matrix it formally follows
 \begin{align}
 \label{formal1}
 &\lim_{t\to\infty }E(-t/2)=SES^{\dagger}=\frac{1}{2}(H_L-H_R),\\
 \label{formal2}
 &\lim_{t\to\infty}E(t/2)=S^{*}E[S^*]^{\dagger}=\frac{1}{2}(H_R-H_L).
 \end{align}
Substituting these relations into the exponents of \eqref{Levitov_def} we obtain \eqref{derivativeJ}.
Notice however that the operator $E(t)$ is non-local and thus the limits taken in (\ref{formal1}, \ref{formal2}) are formal: 
both sides are in fact infinite, and only local observables reach a steady regime. 

\section{Conclusions}
In this paper, we derived the non-equilibrium density matrix for two Ising chains at different temperatures coupled in the middle at time $t=0$, such that the final system is homogeneous.
The density matrix shows a factorized form in terms of right and left moving fermions and agrees with that derived previously using different techniques\cite{aschbacher2003non}. The non-equilibrium steady state supports a time-independent current flowing along the chain, signaling a ballistic mechanism for the heat transport and in agreement
with recent numerical simulations\cite{karrasch2012nonequilibrium}. We have also evaluated for the first time the exact large-deviation function for heat transport in this model.
Although the result for the density matrix is obtained for a free system, it is likely that it
can be extended to a more general class of $1d$ quantum chain models, like integrable models\footnote{B. Doyon, F. Essler and J. E. Moore, work in progress.}.
This is because conserved charges can protect transport from back-scattering, as observed in \cite{sirker2009diffusion}.
Moreover, it may be possible that as in the equilibrium case,
the low-energy physics is universal and well described by a suitable CFT or more generally QFT, where the exact thermal-flow stationary states have already been described (and are similarly factorized)\cite{bernard2012energy,bernard2013non,doyon2012nonequilibrium}.  This would suggest that interactions which are irrelevant at equilibrium are still not relevant when $\beta_l \not= \beta_r$. 

We plan to extend the analysis to the case where the two chains are coupled with
an impurity or more generally when the final Hamiltonian is no more translation invariant \cite{mintchev2013luttinger}. 
A CFT description of this problem is still lacking and it is not even clear whether the steady state will still be factorized or how the presence of the impurity will affect the
current and its fluctuations.
\section*{Appendix}
In this Appendix we give more details on the diagonalization technique of section \ref{solution}. We have\cite{lieb1961two}
\begin{align}
\label{Majorana1}
&\omega_{r/l}^i(k)=\frac{1}{2}[A_{r/l}^i(k)+B_{r/l}^i(k)],\\
\label{Majorana2}
&\xi_{r/l}^i(k)=\frac{1}{2}[A_{r/l}^i(k)-B_{r/l}^i(k)]
\end{align}	
where for the right chain ($1\leq i\leq N$)
\begin{align}
\label{A}
&A_{r}^i(k)=\mathcal{N}_k(-1)^{i-1}\{\sin(i\theta_k)-h^{-1}\sin[(i-1)\theta_k]\},\\
\label{B}
&B_{r}^i(k)=\mathcal{N}'_k(-1)^{i}\sin(i\theta_k).
\end{align}                
The constants $\mathcal{N}_k$ and $\mathcal{N}'_k$ ensures normalization,
e.g. $\sum_{i}A_{r}^i(k)A_{r}^i(k')=\delta_{k,k'}$ and $\sum_k A_{r}^i(k)A_r^j(k)=\delta_{i,j}$.
Up to such normalizations, analogous functions in the left chain are given by ($-N+1\leq i\leq 0$)
\beq
\label{AB}
A_{l}^i(k)= B^{1-i}_r(k)\quad \text{and}\quad B_{l}^i(k)= A^{1-i}_r(k).
\eeq
In order to take the thermodynamic limit, it is useful to write the 
finite size approximation valid for $h>1$, for the $k$-th solution of \eqref{thetatrasc}
\beq
\label{thetaexp}
 \theta_k = x_k + \frac{f(x_k)}{N+1} + \Ord{\frac{1}{N+1}}^2 
\eeq
where $x_k = \frac{\pi k}{N+1}$ and $f(x) = \arctan\left(\frac{\sin x}{\cos x - h}\right)$.
Substituting \eqref{thetaexp} into (\ref{A} ,\ref{B})  we obtain for $\theta\in[0,\pi]$
\begin{align}
\label{phiplusfinite}
& A_r^{i}(\theta)=\sqrt{\frac{2}{\pi}}(-1)^{i-1} \sin\bigl[i \theta - f(\theta)\bigr],\\
\label{phiminusfinite}
& B_r^i(\theta)=\sqrt{\frac{2}{\pi}}(-1)^{i}\sin(i\theta),
\end{align}
that are properly normalized. For $i\leq 0$ one can use \eqref{AB}. The thermodynamic
limit in the full chain is  obtained taking (\ref{A}, \ref{B}),
and sending $N\to 2N$ and $i \to i-N$. However this time, using \eqref{thetaexp}, we notice
that the solution at the same $\theta$ of \eqref{thetatrasc} splits into two degenerate cases (corresponding to even and odd $k$)
\begin{align}
\label{full1}
&A_1^i(\theta)= \sqrt{\frac{1}{\pi}} (-1)^{i-1}\sin\left(i\theta - \frac{f(\theta)+\theta}{2}\right),\\
\label{full2}
&A_2^i(\theta)= \sqrt{\frac{1}{\pi}} (-1)^{i-1}\cos\left(i\theta - \frac{f(\theta)+\theta}{2}\right),\\
\label{full3}
&B_1^i(\theta)= \sqrt{\frac{1}{\pi}} (-1)^i\sin\left(i\theta + \frac{f(\theta)-\theta}{2}\right),\\
\label{full4}
&B_2^i(\theta)= \sqrt{\frac{1}{\pi}} (-1)^i\cos\left(i\theta + \frac{f(\theta)-\theta}{2}\right).
\end{align}
As expected, the set of functions \eqref{full1}-\eqref{full4} is apart from an irrelevant phase the same one would have obtained starting with periodic
boundary conditions. Two fermionic operators $\Psi_{1}(\theta)$ and $\Psi_{2}(\theta)$ can be then introduced from $A_{1,2}^i$ and $B_{1,2}^i$, in the same way as
$\Phi_{l}(\theta)$ and $\Phi_r(\theta)$ are expanded in terms of $A_{l/r}^i$ and $B_{l/r}^i$,
see \eqref{Bogoliubov} and (\ref{Majorana1}, \ref{Majorana2});  left and right moving fermions in \eqref{RLmovers} are linear
combinations of them
\beq
\left[\begin{array}{cc}
\iR(\theta) \\
\iL(\theta)
\end{array}\right]=\frac{e^{- i  \frac{f(\theta) - \theta}{2}}}{\sqrt 2}\left[\begin{array}{cc} 
            i & -1\\ -i & -1
           \end{array}\right]\left[\begin{array}{cc}
\Psi_1(\theta) \\
\Psi_2(\theta)
\end{array}\right].
\eeq
\bibliography{noneq}{}
\end{document}